\documentclass[twocolumn,preprintnumbers]{revtex4}

\usepackage{graphicx}
\usepackage{epsf}
\usepackage{amsmath,amssymb}
\usepackage{color}
\newcommand{\bvec}{\boldsymbol}
\newcommand{\C}{^{14}\textrm{C}}

\begin{document}
\preprint{KUNS-2772, NITEP 32}
\title{Cluster structures and monopole transitions of $^{14}$C}
\author{Yoshiko Kanada-En'yo}
\affiliation{Department of Physics, Kyoto University, Kyoto 606-8502, Japan}
\author{Kazuyuki Ogata} 
\affiliation{Research Center for Nuclear Physics (RCNP), Osaka University,
  Ibaraki 567-0047, Japan}
\affiliation{Department of Physics, Osaka City University, Osaka 558-8585,
  Japan}
\affiliation{
Nambu Yoichiro Institute of Theoretical and Experimental Physics (NITEP),
   Osaka City University, Osaka 558-8585, Japan}
\begin{abstract}
Cluster structures of $^{14}$C were investigated with a 
method of antisymmetrized molecular dynamics (AMD) combined with 
a $3\alpha+nn$ cluster model while focusing on the monopole excitations and 
linear-chain $3\alpha$ band. 
Variation after parity and angular momentum projections
was performed in the AMD framework, and the generator coordinate method was
applied to take into account various $3\alpha+nn$ cluster configurations. 
Energy spectra and monopole and $E2$ transition strengths of $0^+$, $2^+$, and $4^+$ states 
 were calculated to assign band structures. 
The $0^+_3$ state with remarkable monopole transition was obtained as a vibrational mode of 
the triangle $3\alpha$ configuration. In addition, 
the linear-chain $3\alpha$ band from the band-head $0^+_4$ state was obtained.
$^{10}$Be$+\alpha$ decay widths of $0^+$, $2^+$, and $4^+$ states were evaluated.
$\alpha$ inelastic scattering off $^{14}$C was also investigated by the microscopic coupled-channel 
calculation with the $g$-matrix folding model to propose possible observation of the  $0^+_3$ state
via $\alpha$ scattering experiments.
\end{abstract}
\maketitle

\section{Introduction}
$3\alpha$ clustering is one of the hot subjects of C isotopes. 
A variety of $3\alpha$ cluster structures in $^{12}$C have been 
investigated (see for examples, review articles of Refs.~\cite{Fujiwara80,Yamada:2011bi,Horiuchi:2012,Freer:2014qoa,Funaki:2015uya,Freer:2017gip} and references therein). 
Particularly, the $0^+_2$ state of $^{12}$C has been attracting 
a great interest and discussed as a gas state of $\alpha$ particles.
The strong monopole transition from the ground state 
supports the developed $3\alpha$ structure of the $0^+_2$ state. 
In higher energies than the $0^+_2$ state, 
other kinds of $3\alpha$-cluster structure such as 
a triangle, bending-chain, higher-nodal states have been theoretically suggested
\cite{kamimura-RGM1,uegaki1,uegaki3,Kamimura:1981oxj,Descouvemont:1987zzb,KanadaEn'yo:1998rf,Tohsaki:2001an,Funaki:2003af,Neff:2003ib,
Fedotov:2004nz,Kurokawa:2004ejb,Kurokawa:2005ax,Filikhin:2005nc,Funaki:2005pa,KanadaEn'yo:2006ze,
Arai:2006bt,Chernykh:2007zz,ohtsubo13,Epelbaum:2012qn,Dreyfuss:2012us,Ishikawa:2014mza,Suhara:2014wua,Funaki:2014tda}.

For neutron-rich C isotopes, further rich phenomena of the $3\alpha$ clustering are expected to appear 
because of surrounding excess neutrons. 
One of fascinating topics is the linear-chain $3\alpha$ structure in neutron-rich C.
The linear-chain $3\alpha$ clustering in $^{12}$C was originally proposed by Morinaga
\cite{Morinaga1956,Morinaga1966}, but it is considered to be unstable against bending motion. 
However in the case of neutron-rich C isotopes, the linear-chain clustering can be stabilized owing to 
excess neutrons as predicted in theoretical works
\cite{Itagaki:2001mb,Itagaki:2006ic,Suhara:2010ww,Suhara:2011cc,Maruhn:2010dtc,Baba:2014lsa,Baba:2016sbi,yoshida2016-c14,Baba:2017opd}. In this decade, 
many experiments have been performed to observe cluster states in $^{14}$C
\cite{Soic:2003yg,vonOertzen2004,Price:2007mm,Haigh:2008zz,Freer:2014gza,Fritsch:2016vcq,Yamaguchi:2016oay,Tian:2016vvb}.
One of good tools to search for new cluster states is $\alpha$ resonant scattering, which has been recently 
utilized to study cluster states in such unstable nuclei as $\C$. 
For $^{14}$C, candidate states of the linear-chain band have been reported by 
recent experiments of the $^{10}$Be+$\alpha$ resonant scattering \cite{Freer:2014gza,Fritsch:2016vcq,Yamaguchi:2016oay}, 
but its rotational band members have yet to be confirmed. 

An alternative tool to investigate cluster states is the
$\alpha$ inelastic scattering off target nuclei because developed cluster states 
tend to be 
populated by the $\alpha$ scattering via isoscalar transitions
 \cite{Kawabata:2005ta,KanadaEn'yo:2006bd,Funaki:2006gt,Yamada:2008,Yamada:2011ri,Wakasa:2007zza,Itoh:2011zz,Chiba:2015zxa}.
The $\alpha$ scattering has been 
utilized, in particular, to investigate isoscalar monopole transitions into excited $0^+$ states as done for 
 $^{16}$O to study $0^+$ cluster states \cite{Wakasa:2007zza}.

Our aim is to investigate cluster states in $\C$ focusing on the 
monopole excitations as well as $\alpha$-decay property. In preceding works 
on the linear-chain structure of $\C$ \cite{Suhara:2010ww,Baba:2014lsa,Baba:2017opd}, 
methods of antisymmetrized molecular dynamics (AMD) \cite{KanadaEnyo:1995tb,Kanada-Enyo:2001yji,KanadaEn'yo:2012bj} were applied, but the framework is 
not sufficient to describe detailed $^{10}\textrm{Be}+\alpha$ clustering features. 
The cluster structures of $\C$ have been also studied by 
the generator coordinate method (GCM) of a $3\alpha+nn$ cluster model \cite{yoshida2016-c14}. 
However, the cluster model is not able to properly describe low-energy spectra 
because it is not suitable to describe shell model configurations nor cluster breaking in low-lowing states.
One of the authors \cite{Kanada-Enyo:2014qwn} have studied 
low-lying states of $^{14}$C by  
a calculation of variation after angular momentum and parity projections (VAP) with the AMD model. 
The AMD calculation of VAP reasonably described the 
$0^+$, $2^+$, and $1^+$ spectra and Gamow-Teller transitions from $^{14}$N$(1^+_1)$
except for the anomalously hindered $\beta$ decay of the ground state of $\C$.

In the present paper, we apply the AMD method of VAP combining it with the GCM of the 
$3\alpha+nn$ cluster model, which we call ``VAP+cl-GCM'' in this paper,  
and discuss the energy spectra and band structure of $\C$. A particular attention is paid on 
the monopole excitations and linear-chain band. 
Cluster features of $^{10}$Be+$\alpha$ and $3\alpha+nn$ clusterings are discussed.
The $\alpha$ inelastic scattering off $\C$ are also calculated  
with the microscopic coupled-channel (MCC) calculation using the matter and transition densities 
obtained by VAP+cl-GCM. 
The reaction approach is the 
$g$-matrix folding model, 
where $\alpha$-nucleus coupled-channel potentials are 
microscopically derived by folding the Melbourne $g$-matrix effective nuclear interaction with input 
of densities of the target nucleus from the structure model. 
Similar MCC calculations have been done for study of 
$^{12}$C($\alpha,\alpha')$ and 
 $^{16}$C($\alpha,\alpha')$,
and proved to reproduce cross sections of cluster states
without phenomenologically adjusting parameters of the reaction calculation
\cite{Minomo:2016hgc, Kanada-Enyo:2019prr,Kanada-Enyo:2019qbp}.

The paper is organized as follows. 
The procedure of the structure calculation of VAP+cl-GCM is explained in Sec.~\ref{sec:framework}, and 
structure properties of $\C$ are discussed in Sec.~\ref{sec:results-structure}. The result of the $\alpha$ scattering off $\C$ is shown in
Sec.~ \ref{sec:results-scattering}. 
Finally, a summary is given in Sec.~\ref{sec:summary}.

\section{Formulation of VAP+cl-GCM}  \label{sec:framework}
\subsection{model wave function}
$0^+$ and $2^+$ states of $^{14}$C are calculated with 
the VAP version of AMD combined with the cluster GCM,
in which AMD and $3\alpha+nn$ wave functions are superposed. 

An AMD wave function is given by a Slater determinant of 
single-nucleon Gaussian wave functions as
\begin{eqnarray}
 \Phi_{\rm AMD}({\bvec{Z}}) &=& \frac{1}{\sqrt{A!}} {\cal{A}} \{
  \varphi_1,\varphi_2,...,\varphi_A \},\label{eq:slater}\\
 \varphi_i&=& \phi_{{\bvec{X}}_i}\chi_i\tau_i,\\
 \phi_{{\bvec{X}}_i}({\bvec{r}}_j) & = &  \left(\frac{2\nu}{\pi}\right)^{3/4}
\exp\bigl[-\nu({\bvec{r}}_j-\bvec{X}_i)^2\bigr],
\label{eq:spatial}\\
 \chi_i &=& (\frac{1}{2}+\xi_i)\chi_{\uparrow}
 + (\frac{1}{2}-\xi_i)\chi_{\downarrow}.
\end{eqnarray}
Here ${\cal{A}}$ is the antisymmetrizer, and  $\varphi_i$ is
the $i$th single-particle wave function written by a product of
spatial ($\phi_{{\bvec{X}}_i}$), spin ($\chi_i$), and isospin ($\tau_i$
fixed to be proton or neutron)
wave functions. The width parameter $\nu$ is chosen 
to be the same value $\nu=0.19$ fm$^{-2}$ as Ref.~\cite{Kanada-Enyo:2014qwn}.
The parameters ${\bvec{Z}}\equiv
\{{\bvec{X}}_1,\ldots, {\bvec{X}}_A,\xi_1,\ldots,\xi_A \}$ for 
Gaussian centroid positions and nucleon-spin orientations of 
all nucleons are treated as variational parameters, which are optimized of each state
of $\C$. 
We use five AMD wave functions $(\Phi^{(n)}_\textrm{AMD}$ $n=1,\ldots,5$) 
obtained for five states, 
 $^{14}\textrm{C}(0^+_{1,2},2^+_{1,2},1^+_1)$,  with the energy variation 
after $J^\pi$ (angular momentum and parity) projection in Ref.~\cite{Kanada-Enyo:2014qwn}. 
For more details, the reader is referred to this reference.

\begin{figure}[!h]
\begin{center}
\includegraphics[width=6cm]{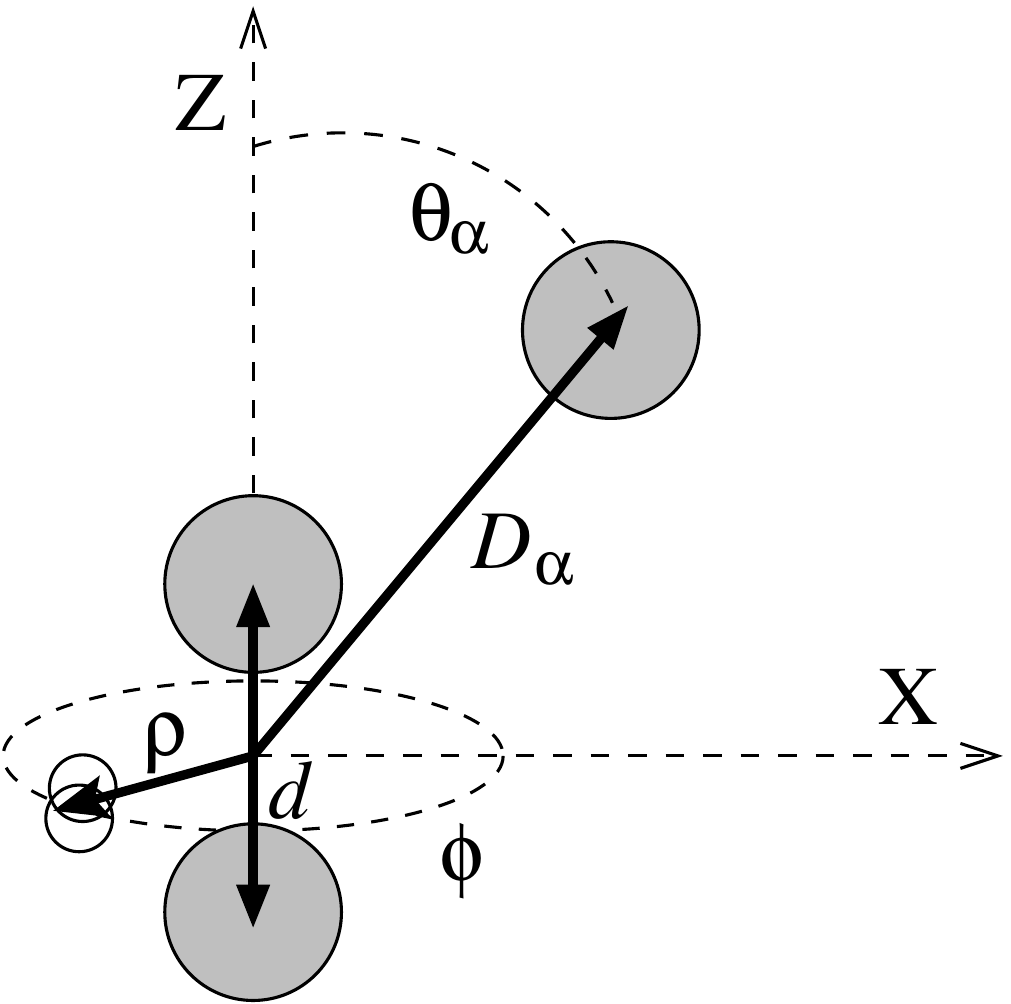} 	
\end{center}
  \caption{Schematic figure of  $(2\alpha+nn)+\alpha$ configurations used 
in the GCM basis wave functions of $^{14}\textrm{C}$.
\label{fig:3-alpha}}
\end{figure}

In addition to the AMD wave functions, 
$3\alpha+nn$ cluster wave functions are superposed to take into 
account large amplitude motion between $\alpha$ clusters. 
The model wave functions for $3\alpha+nn$ configurations are almost 
same as those adopted in Ref.~\cite{yoshida2016-c14}.
We first compose 
wave functions of the subsystem $^{10}$Be with a $2\alpha+nn$ cluster model considering 
the $\alpha$-$\alpha$ motion in $^{10}$Be, and then 
place the third $\alpha$ cluster at various distances and angles from 
the $^{10}$Be core as illustrated by a schematic figure
shown in Fig.~\ref{fig:3-alpha}. 
The first and second $\alpha$ clusters are placed on the $Z$ axis 
at $(0,0,\pm d/2)$ with the distance $d$.
For two valence neutrons, we adopt neutron configurations of
an cluster model for $^{10}$Be proposed by Itagaki {\it et al.} \cite{Itagaki:2005sy},
which is an extended version of the Brink-Bloch cluster model \cite{Brink66}. 
Namely, wave functions for spin-up and -down neutrons  
are given by Gaussian wave packets at the same position but 
the momentum opposite to each other.
As a result, the $2\alpha+nn$ wave function is given 
as 
\begin{align}
&\Phi_{2\alpha+nn}(d,\phi)={\cal A}\Bigl [ 
\Phi_\alpha(\bvec{R}_1)\Phi_\alpha(\bvec{R}_2) \varphi_{n\uparrow}(\bvec{R}_{n\uparrow})
 \varphi_{n\downarrow}(\bvec{R}_{n\downarrow})
\Bigr ], \\
& \varphi_{n\uparrow(\downarrow)}(\bvec{R}_{n\uparrow(\downarrow)})=
\phi_{\bvec{R}_{n\uparrow(\downarrow)}}\chi_{\uparrow(\downarrow)},
\end{align}
with $\bvec{R}_1=-\bvec{R}_2=(0,0,d/2)$ and 
\begin{align}
&\bvec{R}_{n\uparrow}=\rho(\cos\phi,\sin\phi,0)+i\rho\Lambda(-\sin\phi,\cos\phi,0),\nonumber \\
&\bvec{R}_{n\downarrow}=\rho(\cos\phi,\sin\phi,0)-i\rho\Lambda(-\sin\phi,\cos\phi,0).
\end{align}
Here $\Phi_\alpha(\bvec{R}_k)$ is the $\alpha$ cluster wave function 
given by the $(0s)^4$ harmonic oscillator configuration located at $\bvec{R}_k$
and the parameters $\rho$ and $\Lambda$ are optimized to minimize the 
$^{10}$Be energy for each $\alpha$-$\alpha$ distance $d$. 
Using the optimized values of  $\rho$ and $\Lambda$, the $3\alpha+nn$ wave function is given as 
\begin{align}
&\Phi_{3\alpha+nn}(d,D_\alpha, \phi, \theta_\alpha)\\
&={\cal A}\Bigl [ 
\Phi_\alpha(\bvec{R}_1)\Phi_\alpha(\bvec{R}_2) \Phi_\alpha(\bvec{R}_3)
\varphi_{n\uparrow}(\bvec{R}_{n\uparrow})
 \varphi_{n\downarrow}(\bvec{R}_{n\downarrow})
\Bigr ], \\
& \bvec{R}_3=(D_\alpha\cos\theta_\alpha,0, D_\alpha\sin\theta_\alpha).
\end{align}
To exactly remove the center of mass motion from the total wave function, 
the Gaussian center positions are shifted 
as $\bvec{R}_k\to \bvec{R}_k-\bvec{R}_G$ and $\bvec{R}_{n\uparrow(\downarrow)}\to \bvec{R}_{n\uparrow(\downarrow)}
-\bvec{R}_G$ with 
$\bvec{R}_G\equiv (4\sum_k\bvec{R}_k+ \bvec{R}_{n\uparrow}+\bvec{R}_{n\downarrow})/14$.
The Gaussian widths of the $\alpha$ and neutron wave functions are common and 
taken to be the same $\nu$ value as that used in the AMD wave function. 
Note that the present $3\alpha+nn$ cluster wave function can be 
expressed by a specific configuration of the AMD wave function. 

The basis wave functions $\Phi_{3\alpha+nn}(d,D_\alpha, \phi, \theta_\alpha)$
with various values of four parameters, $d$, $D_\alpha$, $\phi$, and $\theta_\alpha$, are projected to 
$J^\pi$ eigen states and superposed in the GCM calculation. We here 
simply denote the $m$th basis wave function as $\Phi_{3\alpha+nn}^{(m)}$ 
with the label $m$ for the parameter set $(d,D_\alpha, \phi, \theta_\alpha)$. 
The final wave function of VAP+cl-GCM for the $J^\pi_k$ state 
is obtained by combining the AMD and the $3\alpha+nn$ wave functions as
\begin{align}
\Psi(J^\pi_k)&=\sum_{n,K} c^{(n)}_{K}(J^\pi_k) P^{J\pi}_{MK} \Phi^{(n)}_\textrm{AMD} \nonumber\\
&+ \sum_{m,K} c'^{(m)}_{K}(J^\pi_k)   P^{J\pi}_{MK} \Phi^{(m)}_{3\alpha+nn},
\end{align}
where coefficients $c^{(n)}_{K}(J^\pi_k)$  and $c'^{(m)}_{K}(J^\pi_k)$ are determined 
by diagonalization of the norm and Hamiltonian matrices. 

In order to evaluate component of a specific $3\alpha+nn$ configuration contained in the $J^\pi_k$ state, 
we define the squared overlap of $\Psi(J^\pi_k)$ with a basis $3\alpha+nn$ wave function as
\begin{align}
{\cal O}^2_{3\alpha+nn}&(d,D_\alpha, \phi, \theta_\alpha)= \nonumber\\
&|\langle \Psi(J^\pi_k) | 
 P^{J\pi}_{MK} \Phi_{3\alpha+nn}(d,D_\alpha, \phi, \theta_\alpha)\rangle|^2
\label{eq:overlap}.
\end{align}

\subsection{Effective interactions and parameter setting}

We use the effective nuclear interactions of the MV1 (case 3) central force \cite{TOHSAKI} supplemented by 
the spin-orbit term of the G3RS force \cite{LS1,LS2}, and the Coulomb force.
The Bartlett, Heisenberg, and Majorana parameters, $b=h=0.125$ and $m=0.62$,
in the MV1 force are adopted, and the strengths $u_{I}=-u_{II}=3000$ MeV 
of the G3RS spin-orbit force are used. These interaction parameters are consistent with 
those used in Ref.~\cite{Kanada-Enyo:2014qwn}
for the study of  $^{14}\textrm{C}$ with the VAP calculation \cite{Kanada-Enyo:2014qwn}. 
It should be commented that we use the MV1 force consisting of the finite-range 2-body and 
zero-range 3-body forces 
instead of the Volkov force \cite{Volkov:1965zz} used in Refs.~\cite{Suhara:2010ww,yoshida2016-c14},
because spectra of low-lying states in $^{14}$C and also those of cluster states in $^{16}$O were reasonably 
reproduced by the MV1 force but not by the Volkov force.

For the generator coordinates $d$, $D_\alpha$, $\phi$, and $\theta_\alpha$
of the GCM calculation, 
we adopt discrete values, $d=2,3,4$~fm, $D_\alpha=2,3,\ldots,7$~fm,
$\phi=\pi/8,3\pi/8, 5\pi/8, 7\pi/8$, and $\theta_\alpha=0, \pi/8, \pi/4, 3\pi/8, \pi/2$ 
of the $3\alpha+nn$ wave functions. 
Note that the angle ranges 
$0\le \phi\le \pi$ and $0\le \theta_\alpha \le \pi/2$ are equivalent to the full ranges
$0\le \phi\le 2\pi$ and $0\le \theta_\alpha \le \pi$ because of the parity and angular momentum 
projections. 
In this parametrization, 
the number of basis $3\alpha+nn$ wave functions  $\Phi^{(m)}_{3\alpha+nn}$
($m=1,\ldots,m_\textrm{max}$) is 
$m_\textrm{max}=(3\times 4 \times 4 + 3) \times 6=306$.

The values of $\rho$ and $\Lambda$ of neutron wave functions 
are listed in Table \ref{tab:be10spe}. 
In the first choice (set-1), we optimize $\rho$ and $\Lambda$ to minimize 
the $0^+$ energy of the subsystem
$P^{0+}_{00}\Phi_{2\alpha+nn}$.  
In the second choice (set-2), 
we use alternative values of $\rho$ and $\Lambda$ for $d=3$ fm 
determined to minimize
the $(J^\pi,K)=(2^+,2)$ energy of $P^{2+}_{M2}\Phi_{2\alpha+nn}$ so as to globally 
optimize energies of $^{10}$Be$(0^+_1)$, $^{10}$Be$(2^+_1)$, 
and $^{10}$Be$(2^+_2)$. As shown in Table \ref{tab:be10spe}, 
the second choice gives a better result of 
the low-energy energy spectra of $^{10}$Be, in particular, the energy of 
$^{10}$Be$(2^+_2)$. Therefore, we adopt parameters 
 $\rho$ and $\Lambda$
of set-2 in the present calculation of $^{14}$C.

\begin{table}[!ht]
\caption{Optimized values of the $\rho$ and $\Lambda$ parameters 
and excitation energies of $^{10}$Be of the first (set-1) and second (set-2) 
of  $2\alpha+nn$ wave functions.
 \label{tab:be10spe}
}
\begin{center}
\begin{tabular}{ccccc}
\hline
   &  \multicolumn{2}{l}{$\rho$~(fm), $\Lambda$}  & \\ 
                          & set-1 & set-2 &  \\
$d=2$ fm & $(0.9,0.56)$  &  $(0.9,0.56)$  & \\
$d=3$ fm & $(1.5,0.38)$ &  $(1.9,0.02)$  & \\
$d=4$ fm & $(2.1,0.24)$ & $(2.1,0.24)$ &\\
 &  \multicolumn{3}{l}{$E_x$ (MeV)}  & \\ 
                         & set(1) & set(2) &  exp \\
$^{10}$B$(2^+_1)$	&2.57 	& 2.52 	& 	3.368\\
$^{10}$B$(2^+_2)$	& 7.64 	& 6.38 	&	5.958\\
\hline
\end{tabular}
\end{center}
\end{table}

\section{Structure of  $^{14}$C}\label{sec:results-structure}

\subsection{Structure properties}

\begin{figure}[!h]
\begin{center}
\includegraphics[width=8.6cm]{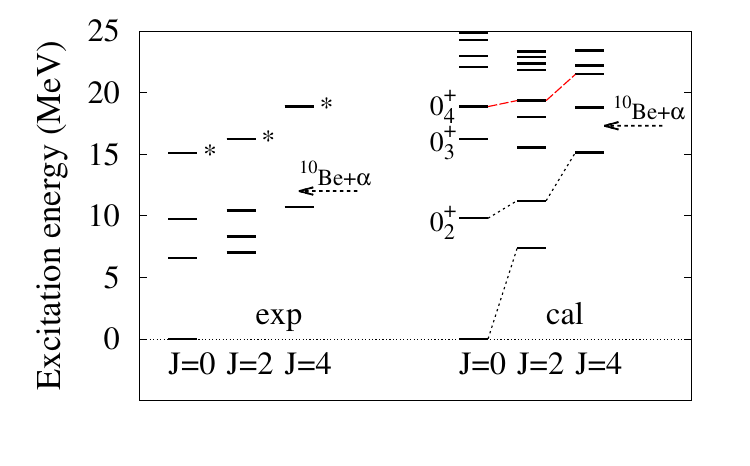}
\end{center}
  \caption{
The calculated energy spectra of 
$0^+$, $2^+$, and $4^+$ states of $\C$ are shown 
together with the experimental energy spectra assigned to $0^+$, $2^+$, and $4^+$ 
in Ref.~\cite{AjzenbergSelove:1991zz}.
The experimental levels with asterisk symbols are the states reported in Ref.~\cite{Yamaguchi:2016oay}
as candidates of the linear-chain band.
  \label{fig:spe}}
\end{figure}

The energy levels of $^{14}$C obtained by the VAP+cl-GCM calculation are 
shown in Fig.~\ref{fig:spe}. 
Root-mean-square radii and monopole transition strengths of the proton and neutron parts for 
$0^+$ states are listed in Table~\ref{tab:radii}, and quadrupole ($\lambda=2$) 
transition strengths of the proton and neutron parts are shown in  Table~\ref{tab:be2}.

\begin{table}[!ht]
\caption{Root-mean-square radii of $^{14}$C$(0^+_k)$
and monopole transition strengths from the ground state  calculated with VAP+cl-GCM. 
Proton and neutron parts of radii  ($R_{p,n}$) and monopole transition strengths 
 ($B_{p,n}(E0)$) are listed. 
The experimental proton radius of the ground state reduced 
from the experimental charge radius \cite{Angeli2013} is $R_p=2.37$~fm.
 \label{tab:radii}
}
\begin{center}
\begin{tabular}{ccccc}
\hline
 &  $R_p$(fm)  & 	$R_n	$ (fm) & $B_p(E0)$ (fm$^4$) & $B_n(E0)$ (fm$^4$) \\
$0^+_1$	&	2.52 	&	2.65 	&		&		\\
$0^+_2$	&	2.62 	&	2.80 	&	1.1 	&	2.5 	\\
$0^+_3$	&	2.98 	&	2.98 	&	18 	&	15 	\\
$0^+_4$	&	3.48 	&	3.37 	&	1.0 	&	0.7 	\\
\hline
\end{tabular}
\end{center}
\end{table}

\begin{table}[!ht]
\caption{$E2$ transition strengths of $2^+\to 0^+$ in 
$^{14}$C calculated with VAP+GCM.
In addition to the proton contribution $B(E2)$, 
the neutron contribution ($B_n(E2)$) of the $\lambda=2$ transitions is also shown.
The strengths for $0^+_1 \to 2^+_{1,2,3,4,5}$ and those with 
$B(E2) \ge 2$ fm$^4$ or $B_n(E2) \ge 2$ fm$^4$ in the 
transitions $2^+_{1,2,3,4,5}\to 0^+_{1,2,3,4}$ are given. 
The experimental value of $B(E2;2^+_1\to 0^+_1)$ is 3.6$\pm$0.6 fm$^4$ \cite{AjzenbergSelove:1991zz}.
 \label{tab:be2}
}
\begin{center}
\begin{tabular}{ccccc}
\hline
 &  $B(E2)$ (fm$^4$)  & $B_n(E2)$ (fm$^4$) \\
$2^+_1\to  0^+_1$	&	7.8 	&	2.4 	\\
$2^+_2\to  0^+_1$	&	0.0 	&	1.5 	\\
$2^+_3\to  0^+_1$	&	0.1 	&	0.0 	\\
$2^+_4\to  0^+_1$	&	0.0 	&	0.0 	\\
$2^+_5\to  0^+_1$	&	0.1 	&	0.1 	\\
&\\			
$2^+_1\to  0^+_2$	&	0.6 	&	2.7 	\\
$2^+_2\to  0^+_2$	&	6.8 	&	43.2 	\\
$2^+_3\to  0^+_2$	&	3.8 	&	3.7 	\\
$2^+_4\to  0^+_1$	&	3.3 	&	0.7 	\\
$2^+_4\to  0^+_3$	&	6.0 	&	2.3 	\\
$2^+_5\to  0^+_3$	&	29.0 	&	33.5 	\\
$2^+_5\to  0^+_4$	&	162 	&	175 	\\
\hline
\end{tabular}
\end{center}
\end{table}

The calculated energy spectra of 
$0^+$, $2^+$, and $4^+$ states are compared with the experimental energy spectra
in Fig.~\ref{fig:spe}. In the figure, 
the experimental levels with asterisk symbols are
candidate states of  the linear-chain band reported by recent
experiment of Ref.~\cite{Yamaguchi:2016oay}.
The $0^+_1$ and $0^+_2$ states are approximately described by the 
VAP wave functions showing shell-model like structure but no prominent 
cluster structure. In the mean-field picture, 
the $0^+_2$ state is approximately understood as 
neutron excitation into a $(sd)^2$ configuration and has a normal proton radius 
as small as that of the ground state. 
In contrast to the $0^+_2$ state, 
the $0^+_3$ and $0^+_4$ states have relatively large nuclear radii and 
spatially developed cluster structures. The $0^+_4$ state dominantly has the linear-chain $3\alpha$ structure and constructs
a rotational band. 

In the monopole transition strengths listed in Table~\ref{tab:radii}, 
the remarkable strength from the ground state is obtained for 
the $0^+_3$ state because of its nature of the triangle $3\alpha$ vibration mode. 
On the other hand, 
the $0^+_4$ state has 
weak monopole transition strength even though it is a developed cluster state, 
because the linear-chain state has the linearly aligned $3\alpha$ configuration, which is 
much different from the 
ground state and difficult to
be directly excited by the monopole operator.

Let us discuss the band structure based on the $E2$ transition strengths shown in Table~\ref{tab:be2}.
The $2^+_1$ state is assigned to the ground band and mainly contributed by the proton rotation. 
The $0^+_2$, $2^+_2$, and $4^+_1$ states with dominant neutron $(sd)^2$ components
construct the $K^\pi=0^+_2$ band.
The $0^+_3$ state shows no clear signal of band structure. Instead, the  
$E2$ strength from this state is fragmented into $2^+_4$ and $2^+_5$ states.
The rotational band of the linear-chain structure is built on the $0^+_4$ state 
with the $2^+_5$ and $4^+_3$ states. 
This band has remarkably strong $E2$ transitions and large moment of inertia (small level spacing) 
because of the highly elongated structure. More detailed properties of $\alpha$ decays and
cluster structures are discussed in the following sections. 

\subsection{$^{10}\textrm{Be}(0^+_1)+\alpha$ decay widths}

\begin{figure}[!h]
\begin{center}
\includegraphics[width=8cm]{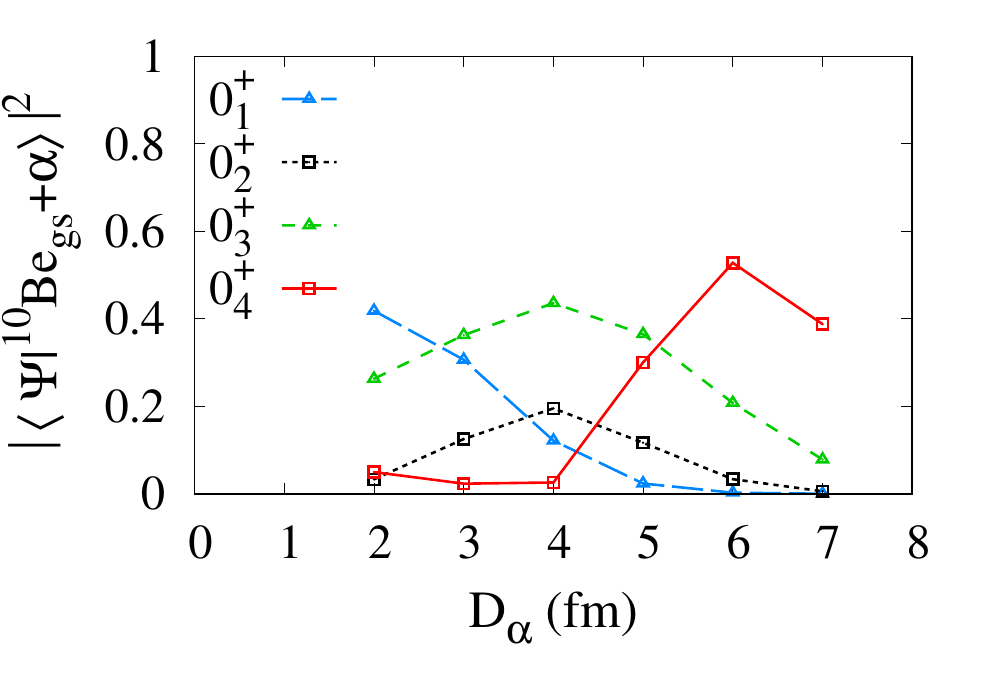}
\end{center}
  \caption{The squared overlap of $\Psi(J^\pi_k)$ for the $0^+_{1,2,3,4}$ states 
with the $^{10}\textrm{Be}(0^+_1)+\alpha$ wave function 
at the distance $D_\alpha$.
  \label{fig:be10-alpha}}
\end{figure}

\begin{figure}[!h]
\begin{center}
\includegraphics[width=9.cm]{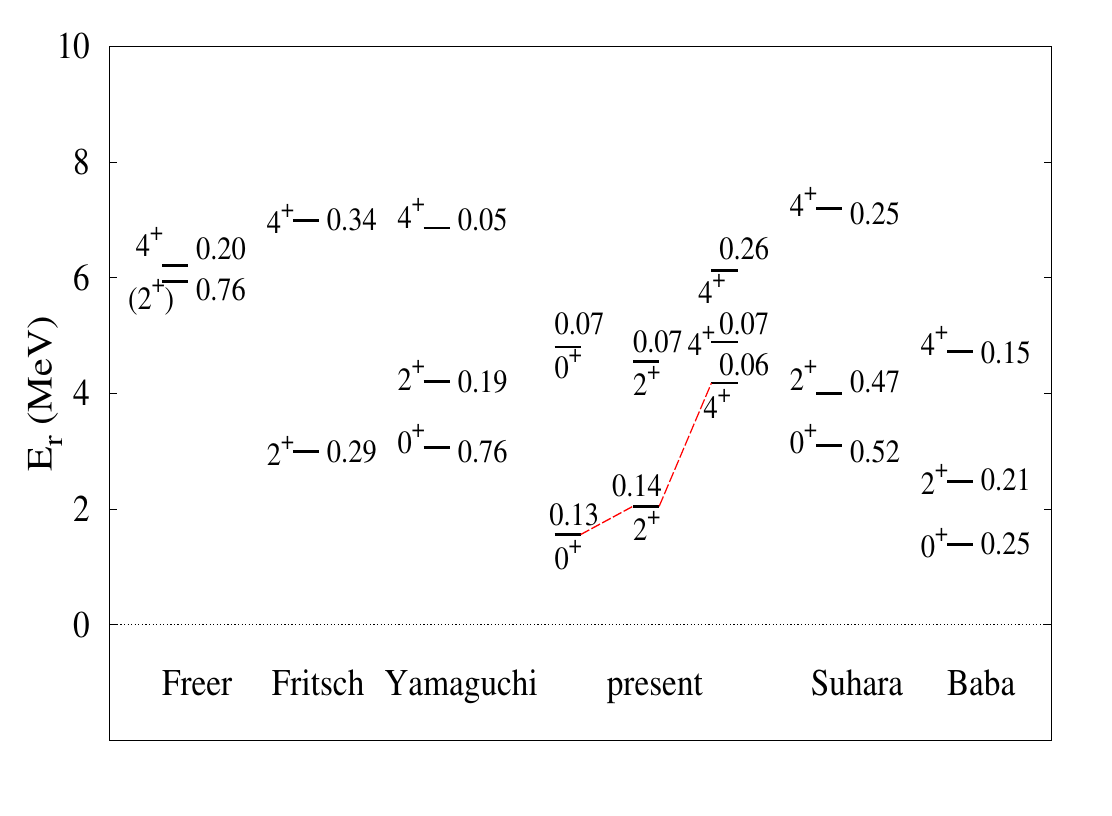}
\end{center}
  \caption{
The calculated energy spectra measured from the $^{10}\textrm{Be}(0^+_1)+\alpha$ threshold 
are shown together with $^{10}\textrm{Be}(0^+_1)+\alpha$ decay widths. 
The partial decay widths are calculated with the reduced widths and penetrability for $E_r$ at the channel radius $a=5$ fm.
The levels connected by dashed lines are the linear-chain band members. 
In the right two columns, the linear-chain band members predicted 
in other calculations by Suhara {\it et al.} \cite{Suhara:2010ww} and Baba {\it et al.} \cite{Baba:2017opd} 
are shown. In the left three columns, 
the $0^+$, $2^+$ and $4^+$  states experimentally observed by $\alpha$ resonant scattering 
in Ref.~\cite{Freer:2014gza} by Freer {\it et al.}, 
Ref.~\cite{Fritsch:2016vcq} by Fritsch: {\it et al.}, and 
Ref.~\cite{Yamaguchi:2016oay} by Yamaguchi {\it et al.}
\label{fig:spe-Er}}
\end{figure}

In order to discuss $\alpha$ decay properties, we calculate the 
$^{10}\textrm{Be}(0^+_1)+\alpha$ component in the $0^+$, $2^+$, and $4^+$ states.
The squared overlap of $\Psi(J^\pi_k)$ obtained by VAP+cl-GCM with 
the $^{10}\textrm{Be}(0^+_1)+\alpha$ wave function with the distance 
$D_\alpha$ is calculated. The result for the $0^+_{1,2,3,4}$ states is 
shown in Fig.~\ref{fig:be10-alpha}. 
The ground state has the overlap 
only in the internal region indicating the 
$^{10}\textrm{Be}(0^+_1)+\alpha$ component as the ground state correlation
but no spatially developed clustering.
In the $0^+_2$ state,  $^{10}\textrm{Be}(0^+_1)+\alpha$ component is relatively minor
compared with the $0^+_3$ and $0^+_4$ states. 
In the  $0^+_3$ state, the squared overlap is 
distributed in a wide range of $D_\alpha$ with the maximum amplitude at 
$D_\alpha=4$~fm meaning that 
the $\alpha$ cluster is moving in the broad region around 
the $^{10}\textrm{Be}(0^+_1)$ core. 
The $0^+_4$ state has large overlap amplitudes at $D_\alpha=6$~fm
and shows remarkable development of the clustering. The significant 
amplitude at  $D_\alpha=7$~fm may indicate some
coupling  with $^{10}\textrm{Be}(0^+_1)+\alpha$  continuum states. 

We evaluate the reduced widths 
for $^{10}\textrm{Be}(0^+_1)+\alpha$ from 
the calculated overlap with the approximation 
proposed in Ref.~\cite{Kanada-Enyo:2014mri}. 
The energy spectra and widths for the states having significant decay widths 
are shown in Fig.~\ref{fig:spe-Er}. 
The energy levels are plotted with respect to relative energies $E_r$ 
measured from the  $^{10}\textrm{Be}+\alpha$ threshold energy. 
Here, the partially decay widths for the $^{10}\textrm{Be}+\alpha$ channel 
are evaluated with the reduced widths and penetrability of $E_r$ at the channel radius $a=5$ fm 
as often done in 
structure model calculations.
The levels connected by dashed lines are the linear-chain band members. 
For comparison, 
theoretical results of the linear-chain bands predicted 
in other two calculations \cite{Suhara:2010ww,Baba:2017opd}.
The theoretical widths are not necessarily consistent between three calculations mainly because of 
different theoretical values of $E_r$, which are affected by model ambiguity from
effective nuclear interactions and model spaces. However, 
the dimensionless reduced widths $\theta(a)^2$ 
are not so much different: 
$\theta(a)^2$ for the $0^+$, $2^+$, and $4^+$ states are
0.15, 0.27, and 0.10 at $a=5$~fm
in the present calculation, 0.16, 0.15, and 0.09 at $a=5$ fm in Ref.~\cite{Suhara:2010ww},
and 0.07, 0.07, and 0.05 at $a=6$~fm in Ref.~\cite{Baba:2017opd}.
It means that the cluster structure of the linear-chain band is qualitatively similar between these 
calculations. 

The data of 
$0^+$, $2^+$ and $4^+$  states reported by experiments of the $\alpha$ resonant scattering 
\cite{Freer:2014gza,Fritsch:2016vcq,Yamaguchi:2016oay} are also shown in left three columns of 
Fig.~\ref{fig:spe-Er}.
It is difficult to 
confirm the experimental assignment of the linear-chain band from those data
because the data are not necessarily consistent between different experiments.
The $\theta_\alpha^2(a)$ values at $a=5$ fm reported 
in Ref.~\cite{Yamaguchi:2016oay} are 0.34(12), 0.091(27), and 0.024(9) 
for the $0^+$, $2^+$, and $4^+$ states, respectively. The observed $\theta_\alpha^2(a)$ values of
the $2^+$ and $4^+$ states are smaller than the present result. 

\subsection{$3\alpha+nn$ cluster structures}

\begin{figure}[!h]
\begin{center}
\includegraphics[width=8cm]{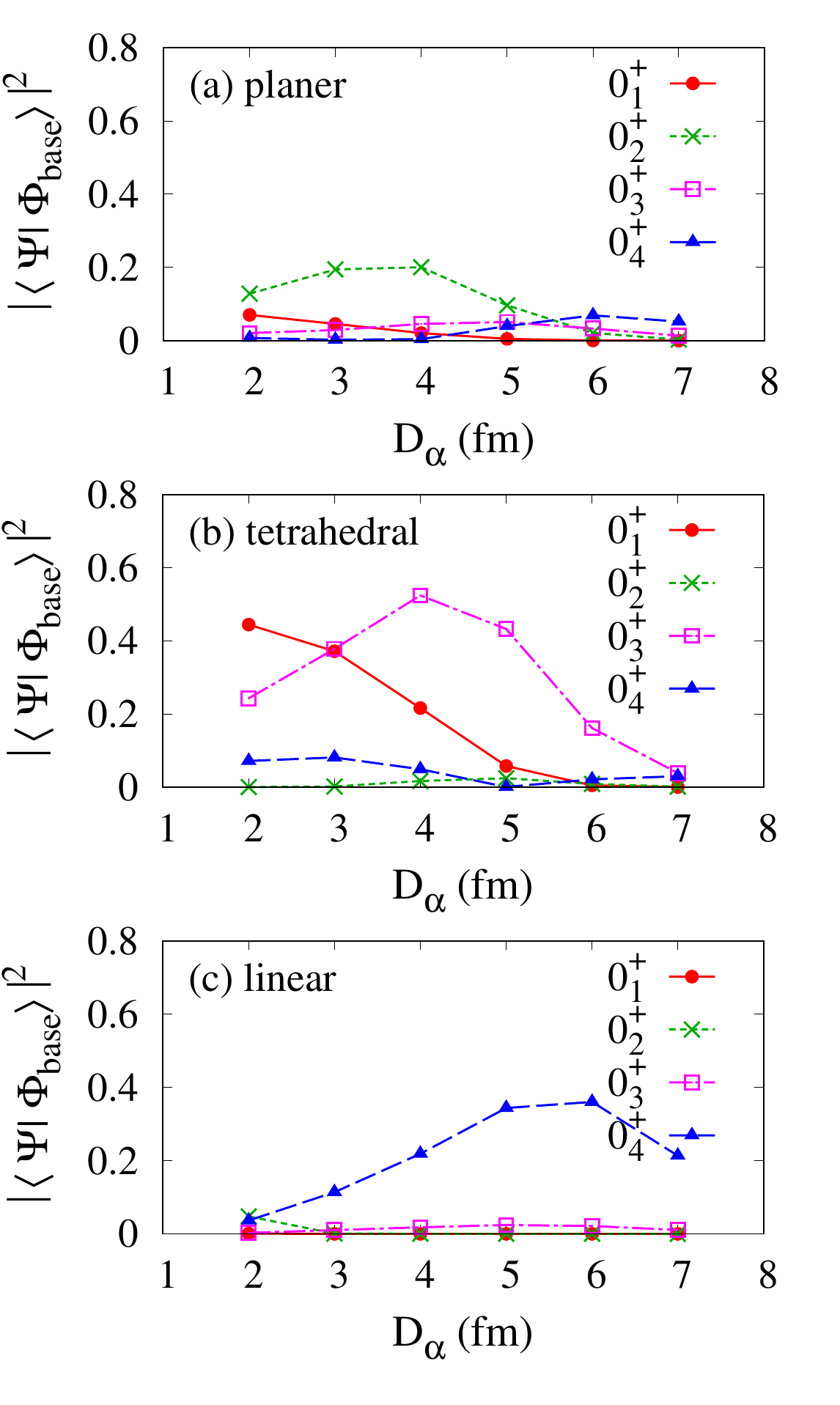}
\end{center}
  \caption{
Components of (a) tetrahedral, (b) planer, and (c) linear configurations contained in $0^+$ states. 
The values of ${\cal O}^2_{3\alpha+nn}(d,D_\alpha, \phi, \theta_\alpha)$ given in Eq.~\eqref{eq:overlap}
with $(\theta_\alpha,\phi)=(\pi/2,5\pi/8)$,  $(\theta_\alpha,\phi)=(\pi/2,\pi/8)$, 
and $\theta=0$ for the tetrahedral, planer, and linear configurations, respectively, 
are plotted as functions of $D_\alpha$. 
$d=4$ fm is chosen for all the configurations.
  \label{fig:over-base}}
\end{figure}

As mentioned previously, 
the remarkable monopole transition strength is obtained for the $0^+_3$ state, 
whereas the strengths of the $0^+_2$ and $0^+_4$ states are relatively weak. These features of 
monopole excitations can be understood by the $3\alpha+nn$ clustering.
In order to clarify properties of the clustering, we calculate 
the squared overlap ${\cal O}^2_{3\alpha+nn}(d,D_\alpha, \phi, \theta_\alpha)$ 
of $\Psi(J^\pi_k)$ with each 
$3\alpha+nn$ configuration specified by the parameters $D_\alpha$, $\theta_\alpha$,
and  $\phi$ as given in Eq.~\eqref{eq:overlap}. 
$D_\alpha$ and $\theta_\alpha$ describes the position of the third $\alpha$
cluster around the $^{10}\textrm{Be}$ core, and
$\phi$ is the parameter for the $nn$ orientation against the $3\alpha$ 
plane (see Fig.~\ref{fig:3-alpha}). 
We categorize $\Phi_{3\alpha+nn}(d,D_\alpha, \phi, \theta_\alpha)$ into three kinds of wave functions 
as ``tetrahedral'', ``planer'', and ``linear'' configurations 
with $(\theta_\alpha,\phi)=(\pi/2,5\pi/8)$,  $(\theta_\alpha,\phi)=(\pi/2,\pi/8)$, 
and  $\theta=0$, respectively. 
Each component contained in $J^\pi_k$ states is evaluated  
with ${\cal O}^2_{3\alpha+nn}(d,D_\alpha, \phi, \theta_\alpha)$ for the corresponding configuration.
Figure \ref{fig:over-base} shows components of the tetrahedral, planer, and linear configurations.
Obtained values of ${\cal O}^2_{3\alpha+nn}(d,D_\alpha, \phi, \theta_\alpha)$ 
are plotted as functions of the distance 
$D_\alpha$ of the third $\alpha$ from the subsystem $2\alpha+nn$. 
Here, we show the result for $d=4$ fm of 
the $\alpha$-$\alpha$ distance in the $2\alpha+nn$ part in order to discuss prominent cluster features.  
As shown in Fig.~\ref{fig:over-base}(a) for the tetrahedral,  
the ground state contains component of the compact
tetrahedral configuration, and the $0^+_3$ state can be regarded as a vibration mode of 
the triangle $3\alpha$ in the tetrahedral configuration. This excitation mode of the $0^+_3$ 
contributes to the remarkable monopole transition strength because it expresses 
radial excitation of $\alpha$ clusters keeping the same shape as the $0^+_1$ state. 
The $0^+_4$ state contains dominantly the linear component with $D_\alpha=5-6$ 
and constructs the linear-chain band with a large momentum of inertia 
because of the highly elongated shape.
The $0^+_2$ state 
shows weak cluster feature as it is the shell-model state and roughly 
described by the neutron excitation to the $sd$ shell. 
However, it contains significant planer component as can be seen in Fig.~\ref{fig:over-base}(a).
From this feature found in the mapping onto the cluster model space, one can interpret the 
$0^+_2$ state as the excitation 
from the compact tetrahedral configuration of the ground state into the planer configuration. 
Such excitations into the planer and linear configurations obtained in the $0^+_2$ and $0^+_4$ states 
involve drastic changes of the geometric structure from the ground state,
and describe suppression of monopole transitions in general.

It should be commented that the squared overlap ${\cal O}_{3\alpha+nn}(d,D_\alpha, \phi, \theta_\alpha)$ 
is not concentrated on a specific configuration but somewhat fragmented. It indicates 
non negligible configuration mixing and also may suggest coupling with continuum states.

In the present calculation of VAP+cl-GCM, the linear-chain band is built on the 
$0^+_4$ state in the $3\alpha+nn$ cluster dynamics. This result is qualitatively similar to that 
of the previous calculation with the $3\alpha+nn$-cluster model \cite{yoshida2016-c14}.
On the other hand, the AMD calculations in Refs.~\cite{Suhara:2010ww,Baba:2014lsa,Baba:2017opd} obtained the linear-chain state
as the $0^+_3$ state. It means that a low-lying monopole excitation was missing 
in the AMD calculations  
because they do not fully take into account $3\alpha+nn$ cluster configurations.

\section{$\alpha$ inelastic scattering off $^{14}\textrm{C}$}\label{sec:results-scattering}

As a probe of cluster states, we investigate the $\alpha$ inelastic scattering off  
$^{14}\textrm{C}$ with the MCC calculation. 
A particular attention is paid on the inelastic scattering of the $0^+_3$ 
having the strong monopole transition, which is expected to be strongly populated by 
the $\alpha$ scattering. For this state, the monopole transition matrix elements 
$M_p(E0)$ and $M_n(E0)$ of the proton and neutron parts 
are predicted to be $M_p(E0)=4.3$ fm$^2$ and 
$M_n(E0)=3.9$ fm$^2$, which are the same order as 
$M_p(E0)=5.48\pm 0.22$ fm$^2$ of 
$^{12}\textrm{C}(0^+_2)$ measured by electron pair
emission \cite{Kelley:2017qgh}.

The $\alpha$ inelastic scattering cross sections of the $0^+$ and $2^+$ states of $^{14}\textrm{C}$ 
are calculated using the matter and transition densities
obtained by the present VAP+cl-GCM  calculation. 
The reaction calculation is in principle the same approach as that of 
our previous works on the $^{12}\textrm{C}(\alpha,\alpha')$ and $^{16}\textrm{O}(\alpha,\alpha')$, which successfully 
reproduced inelastic cross sections of cluster states. 
The $\alpha$-nucleus CC potentials are microscopically derived by folding
the Melbourne $g$-matrix effective $NN$ interaction \cite{Amos:2000} with an $\alpha$ density
and the matter and transition densities of $^{14}$C. The channel-coupling of 
$\lambda=0$ and $\lambda=2$ transitions for the 
$0^+_{1,\ldots,10}$ and $2^+_{1,\ldots,10}$ states theoretically 
obtained by VAP+cl-GCM are taken into account.

\begin{figure}[!h]
\begin{center}
\includegraphics[width=8.6cm]{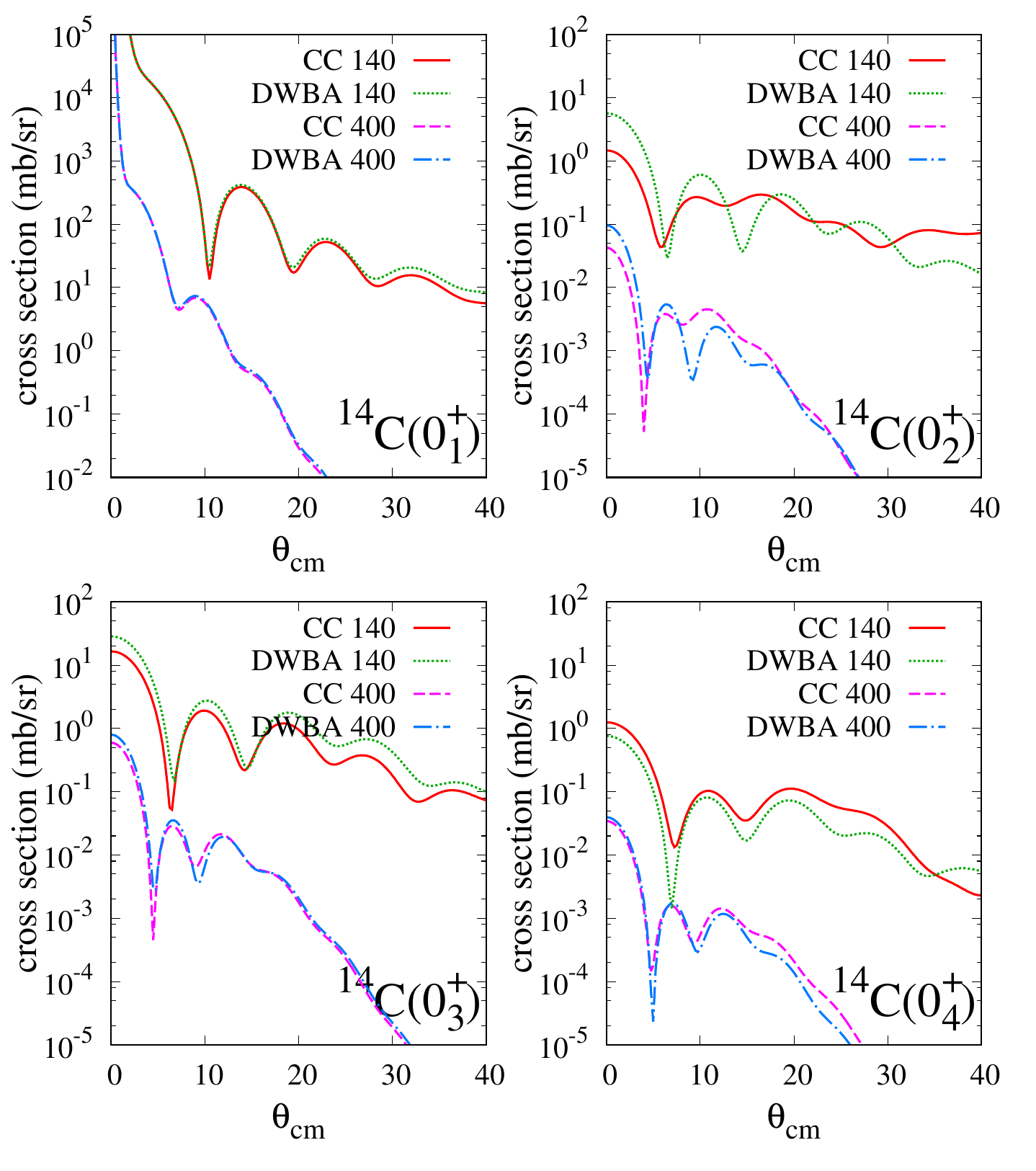}
\end{center}
  \caption{$0^+$ cross sections of the $\alpha$ scattering off $\C$ 
obtained with the MCC and DWBA calculations at incident energies of  
$E_\alpha=140$ MeV and 400 ($\times 10^{-2}$) MeV. 
  \label{fig:cross-c14-1}}
\end{figure}

\begin{figure}[!h]
\begin{center}
\includegraphics[width=8.6cm]{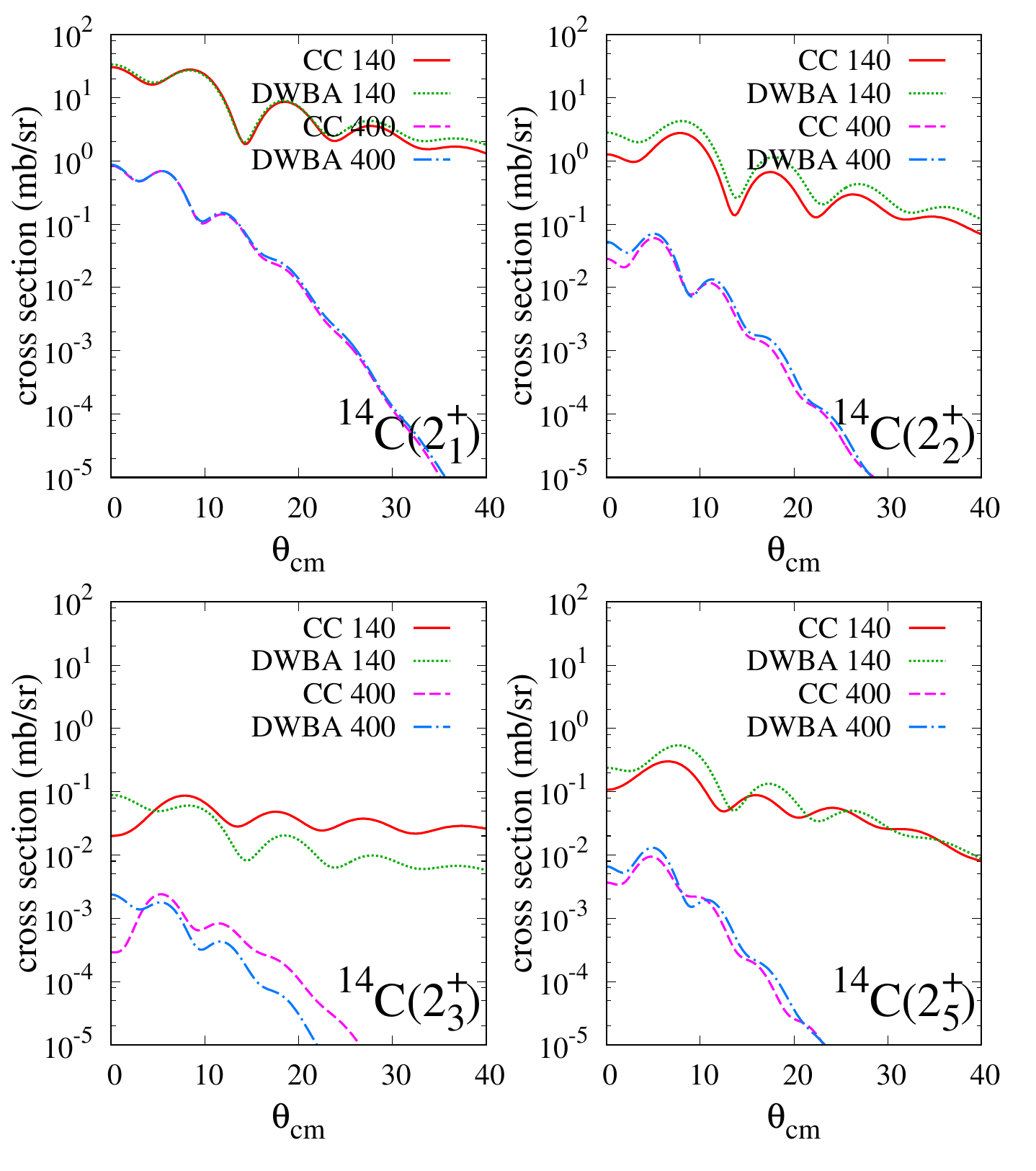}
\end{center}
  \caption{$2^+$ cross sections of the $\alpha$ scattering off $\C$ 
obtained with the MCC and DWBA calculations at incident energies of  
$E_\alpha=140$ MeV and 400  ($\times 10^{-2}$)  MeV. 
  \label{fig:cross-c14-2}}
\end{figure}

The calculated $\alpha$ scattering cross sections of $^{14}\textrm{C}(0^+)$ 
at incident energies of $E_\alpha=140$ and 400 MeV and those of  
$^{14}\textrm{C}(2^+)$ states 
are shown in Figs.~\ref{fig:cross-c14-1} and \ref{fig:cross-c14-2}, respectively.
The strong monopole excitation to the $0^+_3$ state 
by the $\alpha$ scattering is predicted because of the remarkable isoscalar monopole transition.
Inelastic cross sections of the $0^+_4$ state are relatively small. 
This is consistent with the weak isoscalar monopole transitions
from the ground states in prediction of the structure calculation.
Among the  $2^+$ states, 
the cross sections of the $2^+_1$ state in the ground band are significantly large, 
but those of other $2^+$ states are relatively small.

There is no $\alpha$ scattering experiment that observed the $0^+_3$ and $0^+_4$ states. 
From the present prediction of the enhanced $0^+_3$ cross sections, one expects 
strong population of the  $0^+_3$ state in the $\alpha$ scattering. On the other hand, 
the present result suggests relatively weak productions of the linear-chain states, 
the $0^+_4$ and $2^+_5$ states of $\C$, even though they have a developed cluster structure.
However, as seen in Figs.~\ref{fig:cross-c14-1} and \ref{fig:cross-c14-2}, 
the CC effect is minor for the cross sections to those linear-chain states at $E_\alpha=400$ MeV.
This result may suggest a possibility of observing the predicted $0^+_4$
and $2^+_5$ states in the linear-chain band via a multipole
decomposition analysis of the $\alpha$ scattering.

\section{Summary}\label{sec:summary}
Cluster features of $^{14}$C were investigated with VAP+cl-GCM, which is a method of 
AMD combined with the $3\alpha+nn$ cluster model. In the structure calculation of $^{14}$C,  
the AMD wave functions obtained with the variation after parity and angular momentum projections
for $\C(0^+_{1,2})$,   $\C(2^+_{1,2})$, and  $\C(1^+_{1})$ are used. In addition to the AMD wave functions, 
$3\alpha+nn$ wave functions are superposed with  GCM to take into account large amplitude cluster motion.

The energy spectra and band structures of $0^+$, $2^+$, and $4^+$ states were discussed.
The $0^+_3$ and $0^+_4$ states with prominent cluster structures were obtained. 
The $0^+_3$ state has the remarkable monopole transitions from the ground state and 
is regarded as the vibrational mode of the triangle $3\alpha$ configuration.
The $0^+_4$ state contains dominantly the linear-chain $3\alpha$ structure with two neutrons. 
The cluster feature of the $0^+_4$ state is qualitatively consistent with the $0^+_4$ state 
obtained with the $3\alpha+nn$ cluster model calculation in Ref.~\cite{yoshida2016-c14}, 
and similar to the $0^+_3$ state of the AMD 
predictions in Refs.~\cite{Suhara:2010ww,Baba:2014lsa,Baba:2017opd}.

The $^{10}$Be$(0^+_1)$+$\alpha$ components and 
the $\alpha$-decay widths were discussed. 
The calculated result for the linear-chain band was compared with the 
observed data reported by $\alpha$ resonant scattering experiments \cite{Freer:2014gza,Fritsch:2016vcq,Yamaguchi:2016oay}.

The $\alpha$ inelastic scattering off $^{14}$C at $E_\alpha=140$ and 400 MeV 
were also calculated by the MCC
calculation with the Melbourne $g$-matrix interaction for the folding model by
utilizing the matter and transition densities obtained from the structure calculation.
The calculation predicts enhanced monopole cross sections for the $0^+_3$ state
because of the remarkable monopole transition strength, and suggests a possible observation of the 
$0^+_3$ state via an $\alpha$ inelastic scattering experiment in future.
For the $0^+_4$ and $2^+_5$ states of the linear-chain band, the result shows
smaller cross sections 
compared with the $2^+_1$ and $0^+_3$ states. 

In the present calculation, we could not draw a definite assignment of 
the obtained linear-chain band to observed levels 
because model ambiguities remains and also experimental information is still limited.
In the experimental side, many states have been 
observed near and above the  $^{10}$Be+$\alpha$ threshold energy.
The $\alpha$ inelastic scattering can be a good probe for the monopole excitation of 
the triangle vibration mode predicted as the $0^+_3$ state.
For the linear-chain band, the present calculation predicts 
relatively weaker production of the $0^+_4$ and $2^+_5$ states in the $\alpha$ scattering. 
Nevertheless, 
the CC effect is minor for the cross sections to those linear-chain states at $E_\alpha=400$ MeV.
This result may suggest a possibility of observing the predicted linear-chain band via a multipole
decomposition analysis of the $\alpha$ scattering.

\begin{acknowledgments}
The computational calculations of this work were performed by using the
supercomputer in the Yukawa Institute for theoretical physics, Kyoto University. This work was partly supported
by Grants-in-Aid of the Japan Society for the Promotion of Science (Grant Nos. JP18K03617, JP16K05352, and 18H05407) and by the grant for the RCNP joint research project.
\end{acknowledgments}


\begin{thebibliography}{9}


\bibitem{Fujiwara80}
  Y. Fujiwara, Y. Suzuki, H. Horiuchi, K. Ikeda, M. Kamimura, K. Kat\={o}, Y. Suzuki, and E. Uegaki,
  Prog. Theor. Phys. Suppl. {\bf 68}, 29 (1980).


\bibitem{Yamada:2011bi} 
  T.~Yamada, Y.~Funaki, H.~Horiuchi, G.~Roepke, P.~Schuck and A.~Tohsaki,
  Lect.\ Notes Phys.\  {\bf 848}, 229 (2012).


\bibitem{Horiuchi:2012}
H. Horiuchi, K. Ikeda, and K. Kat\=o, Prog.\ Theor.\ Phys.\ Suppl.\ {\bf 192}, 1 (2012).


\bibitem{Freer:2014qoa} 
  M.~Freer and H.~O.~U.~Fynbo,
  Prog.\ Part.\ Nucl.\ Phys.\  {\bf 78}, 1 (2014).

\bibitem{Funaki:2015uya} 
  Y.~Funaki, H.~Horiuchi and A.~Tohsaki,
  Prog.\ Part.\ Nucl.\ Phys.\  {\bf 82}, 78 (2015).

\bibitem{Freer:2017gip} 
  M.~Freer, H.~Horiuchi, Y.~Kanada-En'yo, D.~Lee and U.~G.~Meissner,
  Rev.\ Mod.\ Phys.\  {\bf 90}, 035004 (2018).

\bibitem{kamimura-RGM1}
Y. Fukushima and M. Kamimura,
{\it Proc. Int. Conf. on Nuclear Structure, Tokyo, 1977,
edited by T. Marumori} J. Phys. Soc. Jpn. {\bf 44}, 225 (1978).

\bibitem{uegaki1}
E. Uegaki, S. Okabe, Y. Abe and H. Tanaka, Prog. Theor. Phys. {\bf 57},
1262 (1977).
\bibitem{uegaki3}
E. Uegaki, Y. Abe, S. Okabe and H. Tanaka, Prog. Theor. Phys. {\bf 62}, 1621 (1979).

\bibitem{Kamimura:1981oxj}
  M.~Kamimura,
  Nucl.\ Phys.\ A {\bf 351}, 456 (1981).

\bibitem{Descouvemont:1987zzb}
  P.~Descouvemont and D.~Baye,
  Phys.\ Rev.\ C {\bf 36}, 54 (1987).

\bibitem{KanadaEn'yo:1998rf} 
  Y.~Kanada-En'yo,
  Phys.\ Rev.\ Lett.\  {\bf 81}, 5291 (1998).

\bibitem{Tohsaki:2001an}
  A.~Tohsaki, H.~Horiuchi, P.~Schuck and G.~Ropke,
  Phys.\ Rev.\ Lett.\  {\bf 87}, 192501 (2001).

\bibitem{Funaki:2003af} 
  Y.~Funaki, A.~Tohsaki, H.~Horiuchi, P.~Schuck and G.~Ropke,
  Phys.\ Rev.\ C {\bf 67}, 051306 (2003).

\bibitem{Neff:2003ib} 
  T.~Neff and H.~Feldmeier,
  Nucl.\ Phys.\ A {\bf 738}, 357 (2004).

\bibitem{Fedotov:2004nz} 
  S.~I.~Fedotov, O.~I.~Kartavtsev, V.~I.~Kochkin and A.~V.~Malykh,
  Phys.\ Rev.\ C {\bf 70}, 014006 (2004).

\bibitem{Kurokawa:2004ejb}
  C.~Kurokawa and K.~Kat\=o,
  Nucl.\ Phys.\ A {\bf 738}, 455 (2004).

\bibitem{Kurokawa:2005ax} 
  C.~Kurokawa and K.~Kato,
  Phys.\ Rev.\ C {\bf 71}, 021301 (2005).

\bibitem{Filikhin:2005nc} 
  I.~Filikhin, V.~M.~Suslov and B.~Vlahovic,
  J.\ Phys.\ G {\bf 31}, 1207 (2005).

\bibitem{Funaki:2005pa}
  Y.~Funaki, H.~Horiuchi and A.~Tohsaki,
  Prog.\ Theor.\ Phys.\  {\bf 115}, 115 (2006).

\bibitem{KanadaEn'yo:2006ze} 
  Y.~Kanada-En'yo,
  Prog.\ Theor.\ Phys.\  {\bf 117}, 655 (2007)
  [Erratum-ibid.\  {\bf 121}, 895 (2009)].

\bibitem{Arai:2006bt} 
  K.~Arai,
  Phys.\ Rev.\ C {\bf 74}, 064311 (2006).

\bibitem{Chernykh:2007zz} 
  M.~Chernykh, H.~Feldmeier, T.~Neff, P.~von Neumann-Cosel and A.~Richter,
  Phys.\ Rev.\ Lett.\  {\bf 98}, 032501 (2007).

\bibitem{Epelbaum:2012qn}
  E.~Epelbaum, H.~Krebs, T.~A.~Lahde, D.~Lee and U.~G.~Mei{\ss}ner,
  Phys.\ Rev.\ Lett.\  {\bf 109}, 252501 (2012).

\bibitem{Dreyfuss:2012us}
  A.~C.~Dreyfuss, K.~D.~Launey, T.~Dytrych, J.~P.~Draayer and C.~Bahri,
  Phys.\ Lett.\ B {\bf 727}, 511 (2013).


\bibitem{ohtsubo13}
	S. Ohtsubo, Y. Fukushima, M. Kamimura, and E. Hiyama, Prog. Theor. Exp. Phys. {\bf 2013}, 073D02 (2013).

\bibitem{Ishikawa:2014mza} 
  S.~Ishikawa,
  Phys.\ Rev.\ C {\bf 90}, no. 6, 061604 (2014).

\bibitem{Suhara:2014wua}
  T.~Suhara and Y.~Kanada-En'yo,
  Phys.\ Rev.\ C {\bf 91}, 024315 (2015).


\bibitem{Funaki:2014tda}
  Y.~Funaki,
  Phys.\ Rev.\ C {\bf 92}, no. 2, 021302 (2015).

\bibitem{Morinaga1956} H. Morinaga, Phys. Rev. {\bf 101}, 254 (1956).
\bibitem{Morinaga1966} H. Morinaga, Phys. Lett. {\bf 21}, 78 (1966).

\bibitem{Itagaki:2001mb} 
  N.~Itagaki, S.~Okabe, K.~Ikeda and I.~Tanihata,
  Phys.\ Rev.\ C {\bf 64}, 014301 (2001).


\bibitem{Itagaki:2006ic} 
  N.~Itagaki, W.~v.~Oertzen and S.~Okabe,
  Phys.\ Rev.\ C {\bf 74}, 067304 (2006).

\bibitem{Suhara:2010ww} 
  T.~Suhara and Y.~Kanada-En'yo,
  Phys.\ Rev.\ C {\bf 82}, 044301 (2010).

\bibitem{Suhara:2011cc} 
  T.~Suhara and Y.~Kanada-En'yo,
  Phys.\ Rev.\ C {\bf 84}, 024328 (2011).

\bibitem{Maruhn:2010dtc} 
  J.~A.~Maruhn, N.~Loebl, N.~Itagaki and M.~Kimura,
  Nucl.\ Phys.\ A {\bf 833}, 1 (2010).

\bibitem{Baba:2014lsa} 
  T.~Baba, Y.~Chiba and M.~Kimura,
  Phys.\ Rev.\ C {\bf 90}, no. 6, 064319 (2014).


\bibitem{Baba:2016sbi} 
  T.~Baba and M.~Kimura,
  Phys.\ Rev.\ C {\bf 94}, no. 4, 044303 (2016).

\bibitem{yoshida2016-c14}
Yuta, Yoshida and Yoshiko, Kanada-En'yo,
Prog. Theor. Exp. Phys. {bf 2016}, 123D04 (2016).


\bibitem{Baba:2017opd} 
  T.~Baba and M.~Kimura,
  Phys.\ Rev.\ C {\bf 95}, no. 6, 064318 (2017).

\bibitem{Soic:2003yg} 
  N.~Soic {\it et al.},
  Phys.\ Rev.\ C {\bf 68}, 014321 (2003).

\bibitem{vonOertzen2004}
  W. von Oertzen, H. G. Bohlen, M. Milin, Tz. Kokalova, S. Thummerer, A. Tumino, R. Kalpakpakchieva,
  T. N. Massey,. Y. Eisermann, G. Graw, T. Faestermann, R. Hertenberger, and H.-F. Wirth,
  Eur. Phys. J. A {\bf 21}, 193 (2004).

\bibitem{Price:2007mm} 
  D.~L.~Price {\it et al.},
  Phys.\ Rev.\ C {\bf 75}, 014305 (2007).



\bibitem{Haigh:2008zz} 
  P.~J.~Haigh {\it et al.},
  Phys.\ Rev.\ C {\bf 78}, 014319 (2008).

\bibitem{Freer:2014gza} 
  M.~Freer {\it et al.},
  Phys.\ Rev.\ C {\bf 90}, no. 5, 054324 (2014).

\bibitem{Fritsch:2016vcq} 
  A.~Fritsch {\it et al.},
  Phys.\ Rev.\ C {\bf 93}, no. 1, 014321 (2016).

\bibitem{Yamaguchi:2016oay} 
  H.~Yamaguchi {\it et al.},
  Phys.\ Lett.\ B {\bf 766}, 11 (2017).

\bibitem{Tian:2016vvb} 
  Z.~Y.~Tian {\it et al.},
  Chin.\ Phys.\ C {\bf 40}, no. 11, 111001 (2016).



\bibitem{Kawabata:2005ta} 
  T.~Kawabata, H.~Akimune, H.~Fujita, Y.~Fujita, M.~Fujiwara, K.~Hara, K.~Hatanaka and M.~Itoh {\it et al.},
  Phys.\ Lett.\ B {\bf 646}, 6 (2007).

\bibitem{KanadaEn'yo:2006bd} 
  Y.~Kanada-En'yo,
  Phys.\ Rev.\ C {\bf 75}, 024302 (2007).

\bibitem{Yamada:2008}
T. Yamada1, Y. Funaki, H. Horiuchi, K. Ikeda, and A. Tohsaki, Prog. Theor. Phys. {\bf 120}, 1139 (2008).

\bibitem{Funaki:2006gt}
  Y.~Funaki, A.~Tohsaki, H.~Horiuchi, P.~Schuck and G.~Ropke,
  Eur.\ Phys.\ J.\ A {\bf 28}, 259 (2006).

\bibitem{Wakasa:2007zza} 
  T.~Wakasa, E.~Ihara, K.~Fujita, Y.~Funaki, K.~Hatanaka, H.~Horiuchi, M.~Itoh and J.~Kamiya {\it et al.},
  Phys.\ Lett.\ B {\bf 653}, 173 (2007).

\bibitem{Yamada:2011ri} 
  T.~Yamada, Y.~Funaki, T.~Myo, H.~Horiuchi, K.~Ikeda, G.~Ropke, P.~Schuck and A.~Tohsaki,
  Phys.\ Rev.\ C {\bf 85}, 034315 (2012).

\bibitem{Itoh:2011zz} 
  M.~Itoh {\it et al.},
  Phys.\ Rev.\ C {\bf 84}, 054308 (2011).

\bibitem{Chiba:2015zxa} 
  Y.~Chiba and M.~Kimura,
  Phys.\ Rev.\ C {\bf 91}, no. 6, 061302 (2015).


\bibitem{KanadaEnyo:1995tb}
  Y.~Kanada-En'yo, H.~Horiuchi and A.~Ono,
  Phys.\ Rev.\  C {\bf 52}, 628  (1995).

\bibitem{Kanada-Enyo:2001yji} 
  Y.~Kanada-En'yo and H.~Horiuchi,
  Prog.\ Theor.\ Phys.\ Suppl.\  {\bf 142}, 205 (2001).

\bibitem{KanadaEn'yo:2012bj}
  Y.~Kanada-En'yo, M.~Kimura and A.~Ono,
  PTEP {\bf 2012}  01A202 (2012).

\bibitem{Kanada-Enyo:2014qwn} 
  Y.~Kanada-En'yo and T.~Suhara,
  Phys.\ Rev.\ C {\bf 89}, no. 4, 044313 (2014).

\bibitem{Amos:2000}
K. Amos, P. J. Dortmans, H. V. von Geramb, S. Karataglidis, and J. Raynal,
Adv.~Nucl.~Phys. {\bf 25}, 275 (2000).

\bibitem{Minomo:2016hgc}
  K.~Minomo and K.~Ogata,
  Phys.\ Rev.\ C {\bf 93}, 051601 (2016).

\bibitem{Kanada-Enyo:2019prr} 
  Y.~Kanada-En'yo and K.~Ogata,
  Phys.\ Rev.\ C {\bf 99}, no. 6, 064601 (2019).

\bibitem{Kanada-Enyo:2019qbp} 
  Y.~Kanada-En'yo and K.~Ogata,
  Phys.\ Rev.\ C {\bf 99}, no. 6, 064608 (2019).

\bibitem{Itagaki:2005sy} 
  N.~Itagaki, H.~Masui, M.~Ito and S.~Aoyama,
  Phys.\ Rev.\ C {\bf 71}, 064307 (2005).

\bibitem{Brink66}
	D. M. Brink, {\it Proc. Int. School of Physics Enrico Ferm, Course 36}, Varenna, ed. C. Bloch (Academic Press, New York, 1966).


\bibitem{TOHSAKI}
 T. Ando, K.Ikeda, and A. Tohsaki, Prog. Theor. Phys.
 {\bf 64}, 1608 (1980).
\bibitem{LS1}
 R. Tamagaki, Prog. Theor. Phys. {\bf 39}, 91 (1968).

\bibitem{LS2}
 N. Yamaguchi, T. Kasahara, S. Nagata, and Y. Akaishi,
 Prog. Theor. Phys. {\bf 62}, 1018 (1979).

\bibitem{Volkov:1965zz}
  A.~Volkov,
  Nucl.\ Phys.\  {\bf 74}, 33 (1965).

\bibitem{AjzenbergSelove:1991zz} 
  F.~Ajzenberg-Selove,
  Nucl.\ Phys.\ A {\bf 523}, 1 (1991).


\bibitem{Angeli2013}
I.~Angeli and K.~P.~Marinova, At.~Data Nucl.~Data Tables {\bf 99}, 69 (2013).



\bibitem{Kelley:2017qgh} 
  J.~H.~Kelley, J.~E.~Purcell and C.~G.~Sheu,
  Nucl.\ Phys.\ A {\bf 968}, 71 (2017).


\bibitem{Kanada-Enyo:2014mri} 
  Y.~Kanada-En'yo, T.~Suhara and Y.~Taniguchi,
  PTEP {\bf 2014}, 073D02 (2014).





\end{thebibliography}
\end{document}